\documentclass[11pt]{article}
\usepackage{ieee}
\usepackage{authblk}
\usepackage{comment}
\usepackage{url}
\usepackage{mathtools}
\usepackage{algpseudocode}
\usepackage{algorithm}
\usepackage{algcompatible}
\usepackage{nth}
\usepackage{amssymb}
\usepackage{epsfig}
\usepackage{threeparttable}
\usepackage{footmisc}
\usepackage{hyperref}
\usepackage{paralist}
\usepackage{wrapfig}
\usepackage{enumitem}
\usepackage{subfig}
\usepackage[table]{xcolor}
\usepackage{color}
\usepackage{multirow}
\newcommand{\pubtype}{paper}

\newcommand{\forceindent}{\leavevmode{\parindent=1em\indent}}
\usepackage{graphicx}

\newcommand{\sharma}[2]{\noindent\textcolor{red}{~#1~}\textcolor{blue}{\textbf{(Sharma):}~#2 \\}} 

\begin{document}

\title{Query Processing on Large Graphs: Approaches To Scalability and Response Time Trade Offs}

\author[1]{Soumyava Das\thanks{soumyava.das@teradata.com}}
\author[1]{Abhishek Santra\thanks{abhishek.santra@mavs.uta.edu}}
\author[1]{Jay Bodra\thanks{abhishek.santra@mavs.uta.edu}}
\author[1]{Sharma Chakravarthy\thanks{sharma@cse.uta.edu}}
\affil[1]{IT Lab, CSE Department, University of Texas at Arlington, Texas, USA}


\maketitle

\begin{abstract}
Graphs, being an expressive data structure, have become increasingly important for modeling real-world applications, such as collaboration, different kinds of transactions, social networks, to name a few. With the advent of social networks and the web, the graph sizes have grown too large to fit in main memory precipitating the need for alternative approaches for an efficient, scalable evaluation of queries on graphs of any size.

\forceindent In this ~\pubtype, we use the time-tested ``divide and conquer" approach by partitioning a graph into desired number of partitions (and possibly with appropriate characteristics) and process queries over those partitions to obtain \textit{all or specified number} of answers. This entails correctly computing answers that span multiple partitions or even need the same partition more than once. Given a set of partitions, there are a number of approaches using which a query can be evaluated: i) \textbf{O}ne \textbf{P}artition \textbf{A}t a \textbf{T}ime (\textbf{OPAT}) approach, ii) \textbf{Traditional} use of \textbf{M}ultiple \textbf{P}rocessors (\textbf{TraditionalMP}), and iii) using  the \textbf{Map}/\textbf{Reduce} \textbf{M}ulti-\textbf{P}rocessor approach (\textbf{MapReduceMP}) approach. The first approach, detailed in this paper, has established scalability through independent processing of partitions. The other two approaches address response time in addition to scalability. For the OPAT query evaluation approach, necessary minimal book keeping has been identified and its correctness established in this ~\pubtype. Query answering on partitioned graphs also requires analyzing partitioning schemes for their impact on query processing and determining the number as well as the sequence in which partitions need to be loaded to reduce the response time for processing queries.  We correlate query properties and partition characteristics to reduce query processing time in terms of the resources available. 

\forceindent We also identify a set of quantitative metrics and use them for formulating heuristics to determine the order of loading partitions for efficient query processing. For OPAT approach, extensive experiments on large graphs (synthetic and real-world) using different partitioning schemes analyze the proposed heuristics on a variety of query types. The other two approaches are fleshed out, analyzed, and contrasted with the OPAT approach. An existing graph querying system has been extended to evaluate queries on partitioned graphs. Finally all three approaches are compared for their strengths and weaknesses.
\end{abstract}

\section{Motivation}
\label{sec:introduction}

Querying transactional data stored in a database or searching of documents/web pages is well-established. Search engines do a very good job of retrieving all or top k answers from data repositories. Lately, large data sets are being created (e.g., social networks, web graphs, question/answer graphs, and other ontology-based data sets) that have structural relationships in addition to a wide variety of meta data (multiple labels on nodes and edges, weights on nodes and edges) as part of that structural description. This, in essence, provides an opportunity as well as a challenge to query very large graphs for obtaining insights into the data set as well as efficiently extract desired information (subgraphs) using an expressive query language. Currently, limited search is possible on graphs which is different from answering expressive queries (containing comparison and Boolean operators as well as wild cards.) With the size of the graph databases growing steadily and with the users' need to query (not just mine or search) for complex patterns, expressing them as graph queries and retrieving all or some  answers is very much needed. Hence, approaches for processing queries on large graphs have become a fundamental task (akin to substructure mining, frequent subgraph detection, community and hub detection etc.) for retrieving answers efficiently and effectively with respect to user specified patterns in the form of queries.

\forceindent User specified patterns (as queries) can vary from completely known patterns (where the user exactly knows what s/he is looking for, also termed as an exact match of a subgraph) to include range specification and comparisons, unknown patterns with wild card specification on node and edge labels as well as edge distances in a query. Moreover, the user may also be interested in uncovering results which are combination of multiple patterns (using Boolean operators.) Similar to database querying, this calls for operators (both comparison and logical) to be available for graph querying to satisfy user requirements. Although several techniques for graph querying have been proposed~\cite{kdd/TongFGE07,jbcb/MongioviNGPFS10,icde/TianP08,pvldb/FanLMWW10,pvldb/ZhangYJ10,tkde/JayaramKLYE15,sigmod/KhanLYGCT11} none of them support (expressive) queries.

\forceindent Most of the extant querying or search approaches on graphs either materialize the graph in main memory or keeps it on disk to be staged into memory as needed. Disk-based approaches are good to deal with graph sizes larger than possible in memory, but introduce difficulties for providing customized buffer management and introduces I/O latency. In contrast, partitioned approaches overcome this by partitioning a graph into desired-sized partitions each of which can fit in memory, but introduces complexity in terms of correctness and efficiency in terms of processing required number of partitions either one-at-a-time or in parallel depending upon resource availability. This approach can benefit from not using the partitions that are not needed for processing queries, if they can be identified. This approach has the potential for identifying the sequence in which partitions need to be loaded for processing (whether one at a time or $p$ at a time) using meta information of  partitions and the queries. This approach can also accommodate different processor characteristics by matching partition sizes to processor's memory and computation capabilities.  Finally, this approach is amenable to parallel processing of partitions in multiple ways as well. The numerous benefits of a partitioned approach have motivated the approaches presented in this~\pubtype.

\forceindent In the presence of partitions, query answering starts from a particular partition (or partitions) based on the query plan chosen and may need additional partitions to evaluate the query. The starting node of a query could be present in multiple partitions necessitating a way to rank the starting partitions and choose one (or top k, if multiple partitions are used in parallel.) What makes this even more challenging is that the answer(s) of a query may span multiple partitions requiring us to \textit{load} different partitions in a proper sequence and if needed, the \textit{same partition} more than once in that sequence. This calls for techniques to analyze the choice of partitioning strategies as well as query/partition characteristics to determine the order of loading partitions. 

\forceindent In the ideal case, we want to load the minimum number of required partitions to answer a query (or a batch of queries.) A required partition is one in which one or more of the query plan node exists. If processing a query using multiple processors, minimizing the number of iterations (an iteration is the processing of one or $p$ partitions in parallel) as well as the total number of partitions loaded in all the iterations can be used as an efficiency measure. The lower bound is loading at most the number of required partitions \textit{only once}. Note that the lower bound cannot be determined for a query \textit{only} from analyzing the partitions and the query plan chosen because whether an answer spans multiple partitions can only be determined at run time. Hence, we propose a number of metrics to base our heuristics for choosing partitions after each iteration  to minimize the total number of partitions loaded. This can also be done for a batch of queries rather than each query. Note that the number of partitions loaded may be greater than the number of distinct partitions needed.

\forceindent This ~\pubtype~ focuses on several aspects of the above problem: i) correctness of query evaluation  for partitioned graphs,  ii) proposing and evaluating heuristics for reducing the number of partitions loaded for evaluating a query to produce all answers, iii) an approach (termed \textbf{TraditionalMP}) using $p$ available processors for parallel query evaluation, iv) another approach (termed \textbf{MapReduceMP}) using Map/Reduce framework for parallel query evaluation, and iv) their comparison. This ~\pubtype~ builds upon an existing main memory QP-Subdue~\cite{dawak/DasGC16} graph processing system and extends it to work on graph partitions (termed \textbf{PGQP} or \textbf{P}artitioned \textbf{G}raph \textbf{Q}uery \textbf{P}rocessor.) QP-Subdue uses a cost model for generating query plans and chooses the minimum cost query plan for execution. As the entire graph is loaded into main memory, all start nodes are expanded independently to obtain all answers. In this~\pubtype, we use two widely-used partitioning strategies and identify several metrics for partition evaluation. Furthermore, the paper uses a mix of query  and partition characteristics to determine initial partition(s) for loading and after each partition execution. The contributions of the paper are:
\begin{itemize}
\item First attempt at processing queries using graph partitions for scalability and response time,
\item Extend a main memory graph query processing algorithm and establish its correctness (Section~\ref{sec:preliminaries}, ~\ref{sec:correctness} and ~\ref{sec:implementation}),
\item Metrics and heuristics using partition information and query characteristics to reduce the number of  partitions loaded for query evaluation (Section~\ref{sec:metrics-heuristics}),
\item Extensive experimental analysis for the OPAT approach that validates  proposed heuristics on real-world and synthetic data sets using a broad range of queries (Section~\ref{sec:experimental-validation}), and
\item Two more approaches (TraditionalMP and MapReduceMP) for parallel evaluation of queries on partitioned graphs are presented from a response time perspective  along with their correctness and analysis (Sections~\ref{sec:approach-2} and ~\ref{sec:approach-3}). Due to space constraints, experimental results could not be included for these approaches.
\end{itemize}

\forceindent Remainder of the~\pubtype~is organized as follows. Section~\ref{sec:related-work} summarizes relevant work. Section~\ref{sec:preliminaries} summarizes query/graph representation, cost-based query plans, and the partitioning approaches used. In Section~\ref{sec:correctness}, we describe the PGQP (Partition-based Graph Query Processor) architecture and correctness. Section~\ref{sec:metrics-heuristics} discusses the metrics computed and proposed heuristics. Section~\ref{sec:implementation}, briefly describes the implementation of PGQP. Section~\ref{sec:experimental-validation} demonstrates the viability of this approach along with the experimental validation of proposed heuristics. 
Section~\ref{sec:approach-2} analyzes parallel evaluation of partitions using k processors (TraditionalMP). Section~\ref{sec:approach-3} introduces Map/Reduce approach to parallel query evaluation (MapReduceMP) and its analysis including comparison of the three approaches. Conclusions are in Section~\ref{sec:conclusions}.
\section{Related Work}
\label{sec:related-work}

\textbf{Searching/Querying:} Querying/searching is useful for retrieving information for understanding the contents of graph databases. The process of finding exact/similar patterns in graphs is a well researched area. In Graph-grep~\cite{graphgrep}, a variable path index based approach is used. A separate hash index is constructed for different path lengths containing all possible paths up to length \textit{l} starting from each node in the graph database. In G-index~\cite{sigmod/YanYH04}, frequent substructures are indexed as a prefix tree by translating the graph into unique edge sequences (called canonical labels) using depth first search (DFS) coding. Using such an index, a search will produce results only if it is frequent. Another technique for query processing, G-Ray \cite{gray}, expands a seed node by finding a matching node followed by bridging both nodes by the best possible path, while proposing a goodness score quantifying the proximity between two nodes that is used to rank the results. The main challenges of the graph query/search techniques discussed above are managing the size of the index in Graph-grep or the size of canonical labels as a prefix tree (G-Index) or information about remaining vertices (G-Ray). The response time increases with respect to the size of the graph. Also, none of the above support a query in its generality. 

\forceindent Recently, there have been some attempts to go beyond search.
QP-Subdue~\cite{dawak/DasGC16} is an attempt to move towards general purpose querying of graphs. It is a main memory approach and has modified a substructure mining algorithm to perform query evaluation. A cost-based plan generator is used. Relational and Boolean operators as well as some wild cards are supported for query specification. Querying by example~\cite{tkde/JayaramKLYE15} is another attempt to work only with RDF tuples. However, it does not handle wild cards and general graph query and is also limited by the size of main memory. This work is an extension of the paper~\cite{Dawak/BodraDSC18} that extends query processing to partitioned graphs.

\textbf{Partitioning Schemes: } Graph partitioning, addressed extensively is to partition a graph into $k$ (roughly equal) partitions. As there are many ways to partition a graph, some metric such as minimizing the number of edges between the partitions (termed cut set) is commonly used. For this work, we use two widely-used systems -- \textbf{METIS} \cite{Metis1995} and Karlsruhe High Quality Partitioning \textbf{(KaHIP)}~\cite{KaHiP2011} -- for partitioning a graph on which queries are processed. 
METIS uses a multilevel algorithm proposed by Chaco \cite{ChacoHendrickson1995}. 
KaHIP implements novel local improvement schemes to fit most kind of graphs such as continental-sized road networks as well as large social networks and web graphs compared to METIS. We have chosen METIS and KaHIP as they are very popular and widely used algorithms with source code availability. Although there are distributed graph databases -- Neo4J is a good example(\url{http://www.neo4j.org/)} -- these are not the same as partitioned graphs and furthermore  they are not directly relevant to the problem being addressed (query plan generation and optimization) in this paper as you have to write programs for computing results. More recent work on graph databases and querying (\url{https://scholarworks.gsu.edu/cs_diss/110/}) addresses path queries using indexing and other techniques which is very different from our problem and approach.

\forceindent Although Map/Reduce has become ubiquitous, to the best of our knowledge, there is no work on using map/reduce framework for query processing using partitions. There have been some work on keyword based graph query answering~\cite{bigdataconf/HaoC0HBH15} and also using XML query processing~\cite{webdb/FegarasLGP11} both of which do not deal with graph structure. Relational query processing (especially join) has been discussed in the map/reduce framework~\cite{sigmod/yangHDPD07}. However, graph partitioning have been used for mining~\cite{dawak/DasC15,tkde/DasC18}. This~\pubtype~ uses several approaches to querying on partitioned approach both from scalability and response time reduction perspective.
\section{Graphs, Queries, Plan Generation, and Partitioning}
\label{sec:preliminaries}

In this~\pubtype, we divide a graph into desirable-sized partitions and process queries by loading the partitions one or $p$ at a time as needed to obtain all or n answers. For partitioning a graph, we use two popular approaches -- METIS and KaHIP.

\begin{figure}[H]
    \subfloat[Input graph with type Information\label{subfig:inputGraph}]{%
        \includegraphics[width=0.32\textwidth]{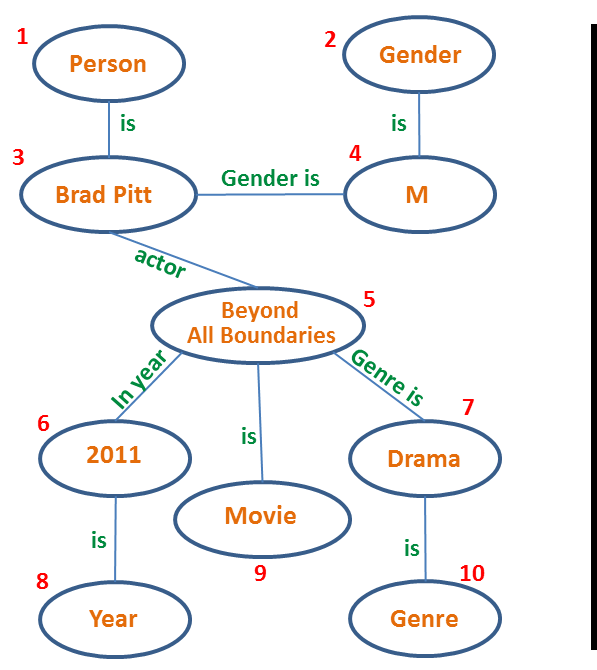}
    }
    \hfill
    \subfloat[Graph partition $P_1$\label{subfig-1:GraphpartitionP1}]{%
        \includegraphics[width=0.28\textwidth]{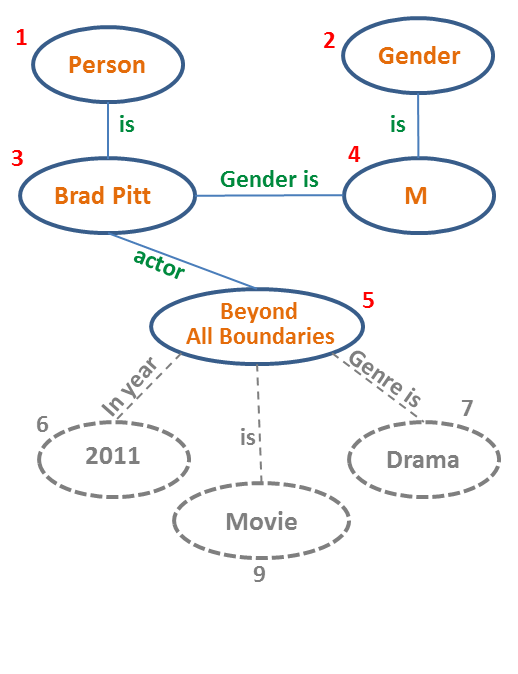}
    }
    \hfill
    \subfloat[Graph partition $P_2$\label{subfig-2:GraphpartitionP2}]{%
        \includegraphics[width=0.28\textwidth]{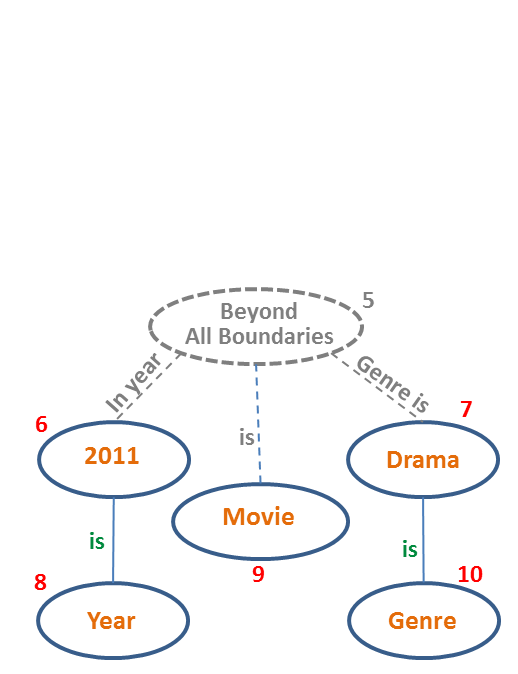}
    }
    \caption{Sample input graph and its partitions}
    \label{fig:Partitions of graph}
\end{figure}

\begin{table}
	\begin{minipage}{0.27\linewidth}
\includegraphics[width=\linewidth]{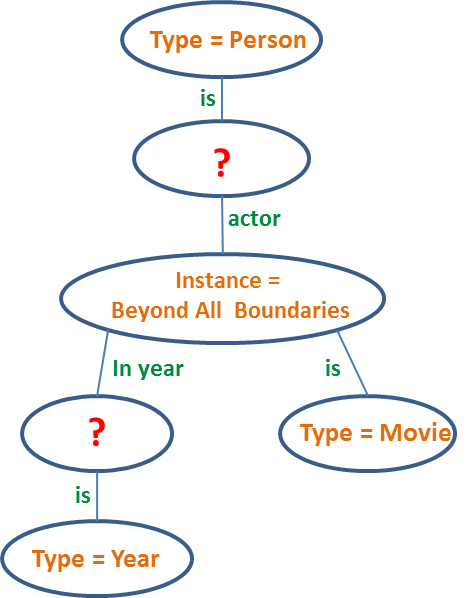}
\captionof{figure}{Representation of a query on input graph in Fig. \ref{subfig:inputGraph}}
\label{fig:queryrep}
	\end{minipage}\hfill
	\begin{minipage}{0.37\linewidth}
		\centering
\small
\begin{tabular}{|l|l|l|}
\hline
\textbf{vID} & \textbf{vL} &  \textbf{pID} \\ \hline
5 & Beyond All Boundaries & 1\tabularnewline
\hline
6 & 2011 & 2\tabularnewline
\hline
7 & Drama & 2\tabularnewline
\hline
8 & Year & 2\tabularnewline
\hline
9 & Movie & 2\tabularnewline
\hline
10 & Genre & 2\tabularnewline
\hline
\end{tabular}
\captionof{table}{Graph representation for partition $P_2$ (Vertex Table)}
\label{tab:vertexTable1}

\end{minipage}\hfill
\begin{minipage}{0.33\linewidth}
\centering
\small
\begin{tabular}{|l|l|l|l|}
\hline
\textbf{dir} & \textbf{s$\_$vID} &  \textbf{d$\_$vID} & \textbf{eL}\\ \hline
u	&	5	&	6	&  In year \\ \hline
u	&	5	&	7	&  Genre is\\ \hline
u	&	5	&	9	&  is \\ \hline
u	&	6	&	8	&  is \\ \hline
u	&	7	&	10	&  is \\ \hline
\end{tabular}
\captionof{table}{Graph representation for partition $P_2$ (Edge Table)}
\label{tab:edgeTable}
\end{minipage}
\end{table}

\noindent\textbf{Graph Representation:} Both undirected and directed graphs can be used for our approach. We follow the Subdue~\cite{subdue2005} representation where vertices are input as an unordered sequence of $<$vertex id (\textbf{vID}), vertex label (\textbf{VL})$>$ pairs and edges are input as unordered tuples of $<$direction (\textbf{dir}), source vertex id (\textbf{s$\_$vID}), destination vertex id (\textbf{d$\_$vID}), edge label (\textbf{eL})$>$. For this approach, a partition number (\textbf{pID}) is added to each vertex in the above representation to keep track of query answers crossing partition boundaries.

\forceindent Figure~\ref{subfig:inputGraph} shows an IMDB (movie database) graph with type information like ``Person", ``Genre", etc. The two graph  partitions $P_1$ and $P_2$ for this graph along with the replicated cut set edges are shown in Figures~\ref{subfig-1:GraphpartitionP1} and~\ref{subfig-2:GraphpartitionP2}, respectively. The broken lines show the additional information kept as part of each partition that corresponds to the cut set.
Tables~\ref{tab:vertexTable1} and~\ref{tab:edgeTable} show the vertex and edge table representation, respectively, for the partitioned IMDB graph shown in Figure~\ref{subfig-2:GraphpartitionP2}. This is used as the graph on which a query is processed.  Figure \ref{fig:queryrep} illustrates the representation of a sample query - ``Find all actors in the movie `Beyond all boundaries' and year of its production".

\noindent\textbf{Cost-Based Plan Generation: } As described in~\cite{dawak/DasGC16,msThesis/Goyal15}, a cost-based plan generation is used along the lines of relational approach to query processing where metadata collected from the database is used to estimate the cost of a query plan, we create a graph catalog which contains information that are relevant to plan generation in a graph database. The graph catalog (generated by making a single pass over the graph database) consists of information, such as type cardinality, average instance cardinality, average connection cardinality, min and max values of type nodes.

\noindent\textbf{Graph Partitioning: } We use METIS and KaHIP  for partitioning and use many of the configurations supported to understand their effect from a query evaluation perspective. Currently, there are no partitioning schemes that generate varying partition sizes (only number of partitions can be specified for METIS ands KaHIP.) In this \pubtype, we have used 2 configurations of METIS - i) kway as partitioning type and sorted heavy edge matching as coarsening type ($kway\_shem$) and ii) recursive bisection as partitioning type and sorted heavy edge matching as coarsening type ($rb\_shem$.) We have used 4 configurations of KaHIP to partition the graph - $fast$, $eco$, $fastsocial$ and $ecosocial$. These 6 strategies are used to partition an input graph into 4 partitions (typically decided based on resource availability. We chose 4 for initial experimentation purpose as our graph sizes did not  warrant larger number.) 
\section{Partitioned Approach to Query Processing}
\label{sec:correctness}
Processing queries on a partitioned graph is very different from processing queries on a single graph. When a graph is partitioned, it will generate \textit{k} graphs $(G_{0},G_{1},...,G_{k})$ such that all the $k$ partitions can be combined to form the original graph. Some additional information is needed (e.g.,  cut set) for combining graphs. If all the answers of a query are in a single partition, it is no different from a non-partitioned approach. However, the case where answers span multiple partitions need to be computed correctly. This entails adding some additional information to each partition to keep track of partial answers that continue into other partitions. These continuations may also require visiting the same partition more than once. This again needs to be handled properly to ensure correctness of results.

\begin{figure}[h]
    \centering
  \includegraphics[width=0.9\textwidth]{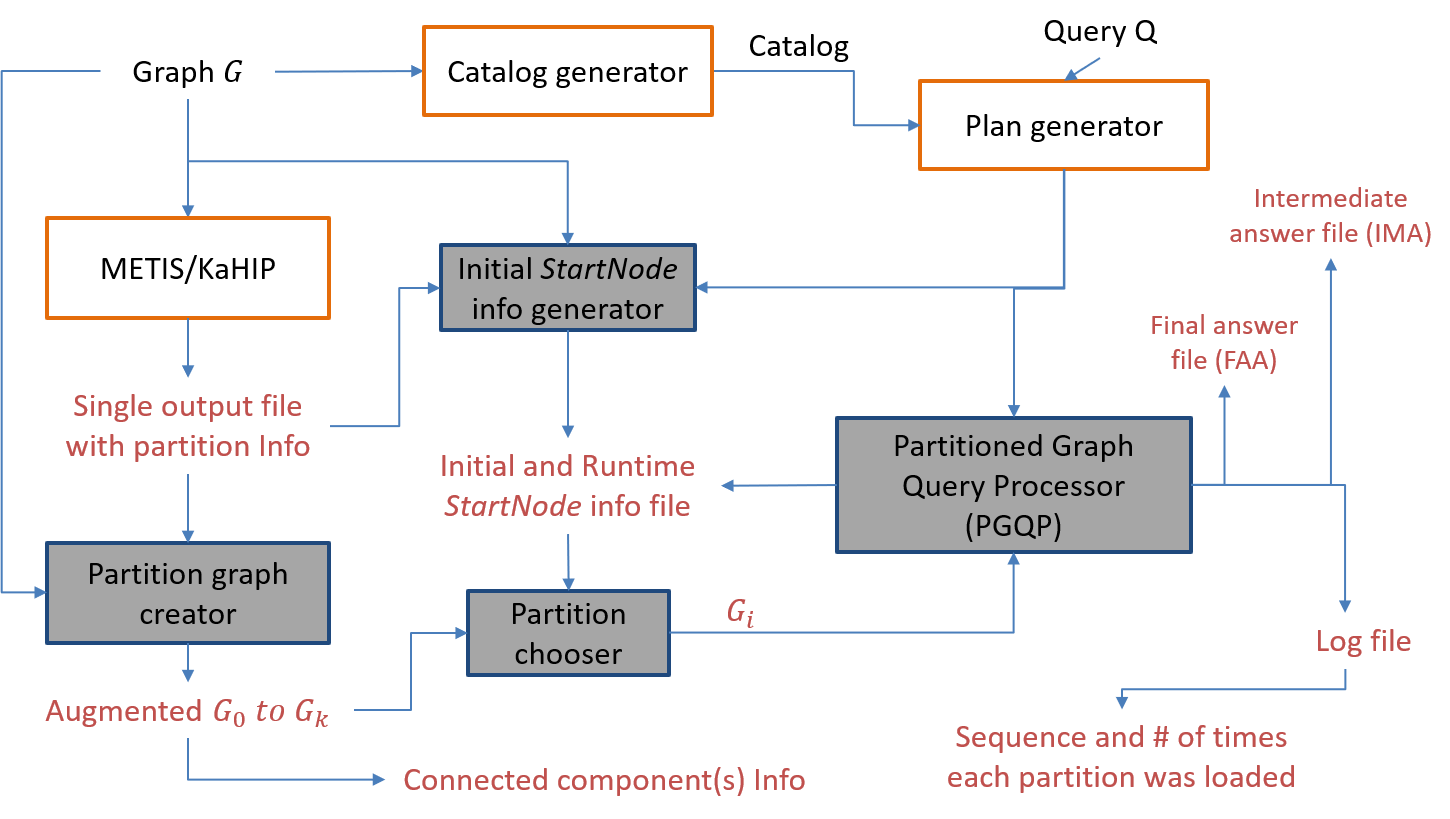}
    \caption{Architecture of the PGQP System}
    \label{fig:SystemArchitecture}
\end{figure}

\forceindent In this~\pubtype, we process three alternative approaches to evaluating queries on partitioned graphs: i) \textbf{O}ne \textbf{P}artition \textbf{A}t a \textbf{T}ime (\textbf{OPAT}) approach, ii) \textbf{Traditional} use of \textbf{M}ultiple \textbf{P}rocessors (\textbf{TraditionalMP}), and iii) \textbf{Map}/\textbf{Reduce} \textbf{M}ulti-\textbf{P}rocessor (\textbf{MapReduceMP}). Our correctness is based on minimal extensions to each partition and keeping track on intermediate results after each partition processing and continuing them as needed.

\subsection{PGQP System Architecture}

Input to the PGQP system is still a graph database and one or more queries. QP-Subdue architecture has been extended (shown in Figure~\ref{fig:SystemArchitecture}) to accept a partition with additional cut set information (instead of the whole graph) and process a query from a set of specified starting points. Our approach uses files for communication between iterations. At the end of an iteration (corresponds to processing one or more partitions independently), each partition appropriate information is written for continuing query processing using other partitions. The shaded modules in Figure~\ref{fig:SystemArchitecture} are extensions for the partitioned approach. The non-shaded modules correspond to pre-existing systems/modules that are used, such as METIS/KaHIP, catalog generator, and plan generator. The partition chooser module chooses the next partition to processing using a specified heuristics or base line as discussed in the rest of the paper.

\forceindent As METIS and KaHIP generate partitions following their own representations, the partition graph creator is implemented to convert the output into desired input for PGQP module as well as compute metrics in one pass. Three metrics are computed by the system: i) number of connected components in each partition (one time), ii) number of start nodes in each partition (one time) and its update after processing a  partition, and iii) sequence and number of times one (or $p$) partition is loaded (at the end) for query evaluation (one time, post-processing.) These are used in various ways as described in the following sections.

\subsection{Correctness of the Approach}
\label{sec:correctness}
\begin{figure}[H]
    \subfloat[Answers within a partition\label{subfig-1:AnswersInSamePartition}]{%
        \includegraphics[width=0.29\textwidth]{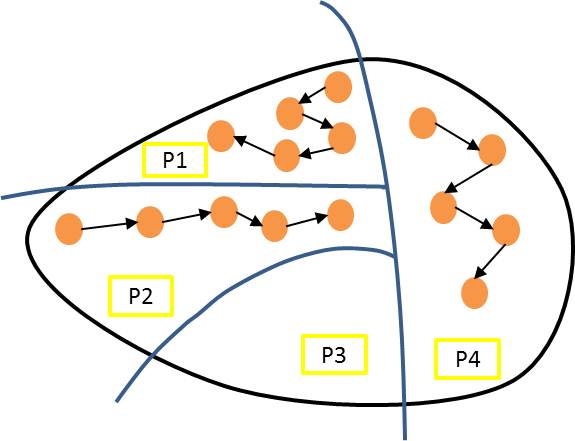}
    }
    \hfill
    \subfloat[Answer crossing multiple partitions\label{subfig-2:AnswersInMultiplePartition}]{%
        \includegraphics[width=0.28\textwidth]{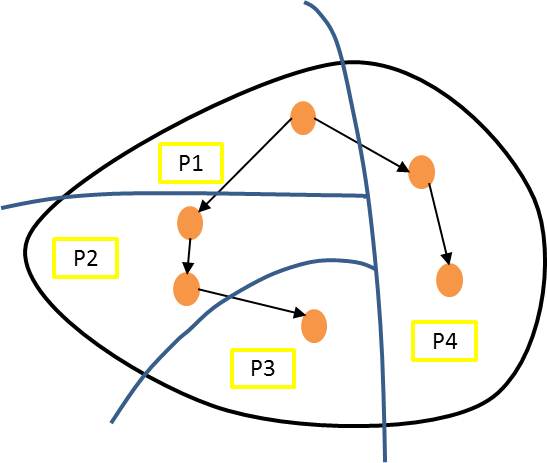}
    }
    \hfill
    \subfloat[Answer using a partition more than once\label{subfig-3:AnswersComeBackInSamePartition}]{%
        \includegraphics[width=0.28\textwidth]{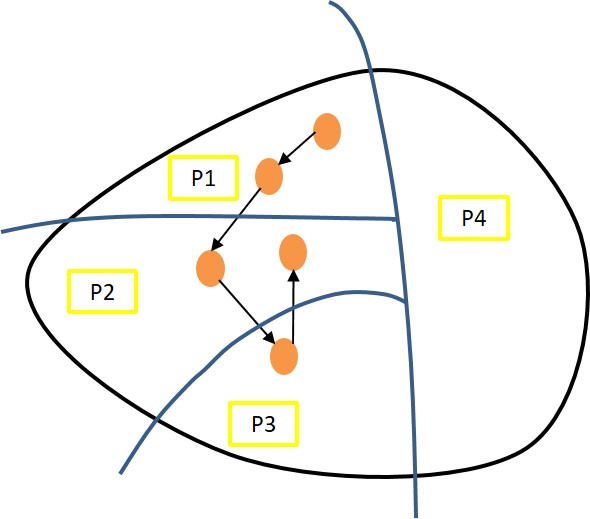}
    }
    \caption{Different Cases of Query Evaluation in Partitioned Graph}
    \label{fig:Partitions of graph}
\end{figure}

The three cases that need to be addressed for answering queries correctly are shown in Figure~\ref{fig:Partitions of graph}. We establish correctness for each case and show how they are implemented in Section~\ref{sec:implementation}. This is common to all the query evaluation algorithms -- OPAT, TraditionalMP, and MapReduceMP --  presented in this~\pubtype. When \textit{all} query results are completely inside a single (or within a) partition as in Figure~\ref{subfig-1:AnswersInSamePartition}, while processing that partition, all results are computed and stored.

\forceindent This is equivalent to QP-Subdue running individually on a (or each) partition as a whole graph. For the case where query results span multiple partitions (illustrated in Figure~\ref{subfig-2:AnswersInMultiplePartition}), when a query expands into another partition, the intermediate answers are written into  a \textbf{P}artition's \textbf{C}ontinuing \textbf{A}nswers (or \textbf{PCA}) file. Importantly, we update the starting node information in the new partitions that the answers span into using the one edge cut set information that has been added to each partition (see Figures~\ref{subfig-1:GraphpartitionP1} and ~\ref{subfig-2:GraphpartitionP2}.) All files are written at the end of processing a partition. When the next partition is loaded (by the partition chooser module based on the heuristic), the \textbf{S}tarting \textbf{N}ode \textbf{I}nformation (or \textbf{SNI}) file (which contains all start nodes in each partition either as labels or as node ids) is used to identify all starting nodes (both start nodes and continuing nodes) for processing. Things that ensure correctness are: i) addition of cut set information to each partition to correctly identify the start nodes in each \textit{continuing partition}, ii) update of SNI file and its usage to determine all start nodes (whether initial or continuation) in that partition, and iii) Carrying and concatenating intermediate results as new partitions are processed.  Figure~\ref{subfig-3:AnswersComeBackInSamePartition} is a special case of the above when an answer instance comes back to a partition that has already been processed.

\forceindent Since a query result can span multiple partitions or even need the same partition more than once, sequencing of partition loads is important from a performance standpoint for all the query evaluation approaches. We discuss the OPAT query evaluation approach first (Section~\ref{sec:metrics-heuristics}.) 
\section{Metrics and Heuristics for Partitioned Query Evaluation}
\label{sec:metrics-heuristics}
One of the widely used metrics for graph partitioning is the \textit{cut set}. Partitioning strategies (e.g., METIS and KaHIP) and partitioning schemes (e.g., kway\_shem of METIS, eco of KaHIP) try to minimize the cut set to reduce inter-partition connections. \textit{In our approach, cut sets are the means by which an answer to a query can span multiple partitions.} In general, if the number of edges in a cut set-cut is small, the likelihood of an answer crossing to another partition from that partition is likely to be small as well. Of course, this depends on the query mix and the characteristics of the \textit{cut  set} (e.g., connecting nodes and labels.) Since our focus is on computing query answers across partitions correctly by loading partitions to maximize efficiency, we want to identify metrics and formulate heuristics to reduce the number of partition loads, the ideal being the minimum required.

\subsection{Number of Query Plan Start/Continuation Nodes in a Partition}
\label{sec:metric1}
For a partitioned approach to query evaluation, we first identify the partitions in which a \textit{query plan start node} exists (termed starting partitions.) Since there can be many start nodes in each starting partition, it is useful to compute the number of such nodes in each starting partition. This is done by using the label information of the start node. It is also possible to rank partitions based on the number of start nodes. If a partition does not have a start node, it may be needed only if an answer spans that partition.  In this \pubtype, we propose two heuristics for choosing a partition to process based on the number of query plan start/continuation nodes in each partition. Note that any query plan node can become a continuing node in a partition during the process of query evaluation. Starting partition for processing a query is determined based on the chosen heuristic. Since these heuristics involve number of query plan start nodes in a partition, \textit{this could change after processing a partition.} Hence, this metric is updated at the end of each partition processing to determine the next partition to be used. This process is continued until there are no more eligible partitions to process. The two heuristics, based on number of start nodes (SN), are discussed below.

\begin{enumerate}
\item \textbf{MAX-SN-heuristic}:
\label{subsec:MAX}
This is similar to the greedy strategy for the choice of the next partition for processing. The partition with the most number of start nodes (from the eligible set of partitions) is chosen. If there is a tie, one is chosen randomly. The intuition behind this heuristic is that highest number of query answers (due to maximum number of start nodes) will be explored  and is likely to help reduce the number of partitions loaded if the answers span into other partitions. Also, if some answers are answered within that partition (initially or while continuing), it will reduce the number of spans. The start node information is updated at end of each partition processing and the process repeated. Note that a previously processed partition may have to be processed again. 

\item \textbf{MIN-SN-heuristic}:
\label{subsec:MIN}
In this case, we load the partition with the least number of start nodes in that partition. Again, ties are resolved randomly. The intuition behind this heuristic is that we accumulate the spanning requirements into partitions with larger number of start nodes in the hope that they can be processed only once.
\end{enumerate}

\forceindent In order to evaluate the proposed heuristics, we need a baseline. We use random way of choosing the initial as well as the next partition from among the set of eligible partitions as our baseline. We completely ignore the number of start nodes in any partition.  We believe that the MAX-SN-heuristic will lead to better performance (in terms of the number of partition loads) than the other heuristic and the baseline. We also believe that either heuristic should do better than the baseline choice (\textbf{RANDOM-SN}.)  When the number of partitions are large, the impact of the proposed heuristics is likely to be more pronounced than when compared to the baseline. These observations have been evaluated for validation in section \ref{subsec:startinglabels}

\subsection{Total Number of Connected Components}
\label{sec:Connected Components in a partition}
Partitioning strategies seem to focus more on the \textit{cut set} and not worry so much about the \textit{number of connected components} generated in each partition. However, for query processing, the number of connected components within each partition and hence the total number of connected components in a partitioning scheme are very important. The partitioning schemes typically produce one partition with a single connected component and the rest of the partitions with varying number of connected components in each of them. We compute the number of connected components in each partition for each partitioning scheme during the same pass in which we compute other metrics.

\forceindent In the presence of disconnected components in a partition, if a query answer spans more than one connected component in that partition, it forces this partition to be \textit{loaded again} for processing. This is avoided (or less likely) if a partition has only one (or small number of) connected component.

\forceindent Instead of analyzing the effect of connected components at the individual partition level, we evaluate the effect of total number of connected components in a partitioning scheme on the performance (i.e., the number of partitions used) of query processing. We believe that \textit{choosing a partitioning scheme with least number of total connected components} (termed \textbf{MIN-CC-heuristic}) is always better from a query evaluation perspective as compared to any other scheme and especially one that produces highest number of total connected components (termed \textbf{MAX-CC-heuristic}). This is likely to make an answer stay inside a partition more often and thereby reduce the total number of partitions used. We will also evaluate this with experiments showing total partition loads using different partitioning schemes of the same graph in section \ref{subsec:connectedcomponents}. Moreover, either of these heuristics will perform better than arbitrarily choosing any of the partitioning schemes randomly as baseline (\textbf{RANDOM-CC}.)

\subsection{Quantitative Measures for Evaluating the Heuristics}
\label{sec:measures}
In the OPAT approach, the total time to answer a query will be directly proportional to the number of partitions loaded \textit{(includes multiple loads of the same partition)} in order to generate all answers. For each query, the ideal case can be inferred from the number of partitions in which a query plan node occurs (actually an upper bound.) We use this as the lower bound on the number of partitions that are needed for processing a query. A quantitative measure for evaluating a heuristic h (e.g., MIN-CC) can be derived by comparing the lower bound ($L_{ideal}$) of partitions to the number of actually loaded partitions ($AL_h$). This load ratio of ideal to actual $({L}_{ideal} / {AL}_{MAX}$) will indicate the effectiveness of a heuristic. This value is at best 1. A heuristic that has a higher value for this ratio is better than the one with a lower value. 

\forceindent For this purpose, we define two measures for MIN-SN and MAX-SN heuristics (denoted by h below) -- one that measures the average load ratio for the \textit{same query} across \textit{partitioning schemes} for a given database D ($\textbf{h(D)}_{pschemes}^{query}$) and another that measures the average load ratio for a \textit{batch of queries} using the \textit{same partitioning scheme} for a given database D ($\textbf{h(D)}_{qbatch}^{pscheme}$). 
For example, $\textbf{MAX-SN(IMDB)}_{pschemes}^{Q1}$ can be expressed as 
$\frac{1}{|pschemes|}{\sum_{pschemes} ({L}_{ideal} / {AL}_{MAX})}{}$, where $AL_{max}$ is the number of actual partitions loaded for Q1 using MAX-SN-heuristic. Others are defined similarly.

\forceindent The second measure defined above is also used for the connected component heuristics: \textbf{MIN-CC-heuristic} and \textbf{MAX-CC-heuristic}. For any of the starting node related heuristics, MIN-CC-heuristic is the average load ratio of a batch of queries executed on the partition scheme with minimum number of connected components. MAX-CC-heuristic is defined similarly. These measures are computed for our experiments to validate our conjecture.

\forceindent Note that CC and SN heuristics are orthogonal or independent of each other. Either can be used individually to improve performance and they can be used in tandem as well for improving the performance further. This provides alternatives to the user depending upon the graph characteristics and the partitioning strategies available.
\section{Implementation Summary}
\label{sec:implementation}

Only the important aspects of implementing the PQGP system (whose architecture is shown in Figure~\ref{fig:SystemArchitecture}) is briefly summarized in this section. For additional details, refer to~\cite{msThesis/Bodra16}. We capture start node and its label information of the query plan for each partition in a \textbf{S}tarting \textbf{N}ode \textbf{I}nformation (\textbf{SNI}) file. This file is updated at the end of processing each partition and used for choosing the next partition based on a heuristic as well as identifying the start nodes in that partition. In addition, we keep an \textbf{I}nter\textbf{m}ediate \textbf{A}nswers (\textbf{IMA}) file -- \textit{one for each partition}. This file stores intermediate or partial query results relevant to that partition. When a query has been answered completely the final results are appended to a \textbf{F}inal \textbf{A}ll \textbf{A}nswers (\textbf{FAA}) file. This file contains all the answers at the end of query processing. Finally, we also keep a \textbf{log file} to compute some run time metrics needed for our analysis.

\subsection{Management of  Partial results}
\label{sec:bookkeeping}

\begin{figure}[h]
   \centering
    \includegraphics[width=\textwidth]{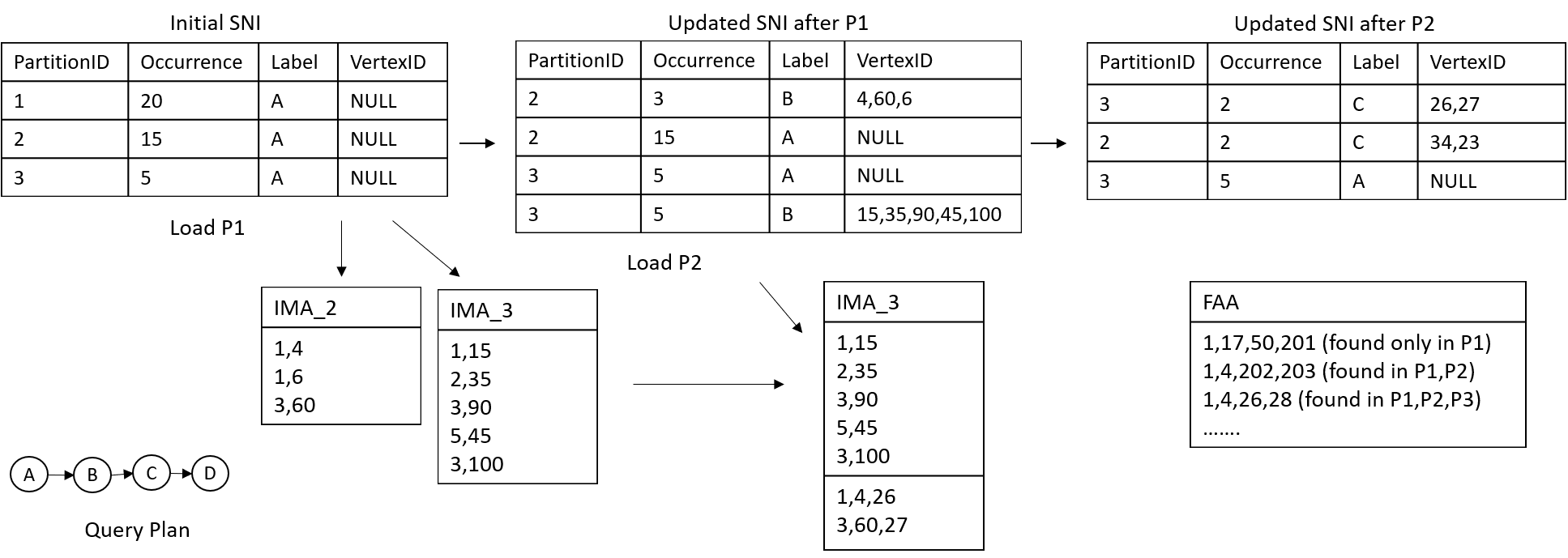}
    \caption{Query answering in the PGQP System}
    \label{fig:exampleDiag}
\end{figure}

For each query, based on the start node label of the query plan, PGQP finds all the relevant partitions containing the start node label, number of occurrences  (or nodes) for that label and generates an initial SNI file. The initial SNI file contains the starting vertex label (according to the query plan) and the number of its occurrences. Figure~\ref{fig:exampleDiag} shows the initial SNI for all partitions with starting node label and number of occurrences. Note that the vertex ids are NULL indicating that they are \textit{start nodes and not continuation nodes}. Continuation nodes will have both a label and a vertex id obtained from the partition extension information.  Following the chosen heuristic (let us assume MAX-SN for this discussion), we pick partition 1 (P1) as our starting partition to process. Expansion process is identical to breadth first search from the starting vertex id and abiding by the query plan.

\forceindent Assume that some answers can be found completely in P1 and some continue in other partitions. At the end of processing P1, complete answers are written into the FAA file (here 1-17-50-201). Final answers can easily be demarcated from partial answers by the size of the answer and edge label(s). Following the query plan, a result having the same size and edge label(s) of the query is a complete answer while any answers with a lesser size is considered a partial result. Partial answers continue in other partitions which are written into corresponding IMA files. For example if 1-4 crossed P1 and moved into P2 this intermediate result is written in $IMA_2$ while the SNI has been updated with the vertex id 4 for B. The SNI after completion of P1 indicates that, query answers can continue from partial answers in P2 and P3 (the ones with vertex ids) or start from node labels A in partitions 2 and 3 (marked with NULL in vertex ids). The vertices already expanded are dropped from the SNI. Continuing with our MAX-SN heuristic we load P2 next as it has 18 vertices in SNI as compared to 10 of P3.

\forceindent Query answers that start in P1 and end in P2 are written to the FAA. See that some query answers continued onto other partitions and the SNI file was updated accordingly. The intermediate results going now into P3 are appended to $IMA_3$. See in Figure~\ref{fig:exampleDiag}, $IMA_3$ is updated by containing both substructures with 2 and 3 vertices. All of these intermediate results will be expanded when P3 is loaded. When all answers are computed, the SNI file will be empty indicating that there are no more partitions to be processed.

\forceindent The SNI file and all IMA files are dropped at the end of query processing. The FAA file contains all the answers added incrementally as soon as an answer is completed. Hence, the user can see results as and when they are generated instead of waiting for the processing to complete. This process can be stopped to generate only k answers.

\forceindent A combination of the SNI, IMA and FAA guarantees correctness. Intuitively SNI contains all starting points or continuation points for partial results while the IMA contains the intermediate or partial results for each partition. After every iteration, the partitions that needs processing have a non-empty IMA file. The SNI file is properly  updated to facilitate easy continuation of the query from the next partition.

\section{Experimental Analysis}
\label{sec:experimental-validation}

Experimental results presented in this section provide an analysis of the proposed heuristics and their validation for the OPAT query processing approach.

\textbf{Experimental Setup:} All experiments have been carried out on Dual Core AMD Opteron 2 GHz processor machine with 16 GB memory. To test the \textit{correctness of our approach}, query results given by QP-Subdue (\cite{dawak/DasGC16}) - a non-partitioned, main-memory query processor for graph databases - have been used as ground truth. The largest graph size we have been able to handle in QP-Subdue on our 16GB machine is 550K nodes and 1700K edges.

\textbf{Datasets:} We have used two datasets: i)
The Internet Movie Database - \textbf{IMDB} graph (1750KV, 5100KE) containing information of movies, actors, genres, year, company, etc. The vertex labels in IMDB are unique, thus the result set for any query is small (some times one.)
ii) A \textbf{synthetic graph} generated using Subgen with 400K vertices, 1200K edges, 2000 unique vertex labels and 4000 unique edge labels with uniform distribution to analyze queries with multiple results.

\textbf{Query Characteristics:} Three different queries were used on IMDB
- Find tv-series and their production companies from the animation \textbf{AND} comedy genres that had ``Kelsey Wagner" as an actor and matched with person is (\textit{Query 1}), List the ``Adam Sandler" movies and their production companies that belonged to the comedy \textbf{AND} Sci-Fi genres but the release year was \textbf{NOT EQUAL} to 2000 (\textit{Query 2}) and, Find all the production companies where ``Fred Wolf" has worked as a writer \textbf{OR} ``Salma Hayek" has worked as an actress (\textit{Query 3}.) These formulated queries have different characteristics that are relevant to the partitioning problem, such as query answers completely inside a single partition (Query 3), query answers spanning multiple partitions for exact results (Query 2), and queries that need to use the same partition more than once (Query 1). Moreover, we have used different combinations of comparison operators ($<$, $<=$, $>$, $>=$, $!=$, $=$) and logical operators (OR, AND). Query plans were generated for all the queries on IMDB data sets using QP-Subdue.

\begin{figure}[htb]
  \centering
  \begin{minipage}[b]{0.36\textwidth}
    \includegraphics[width=\textwidth]{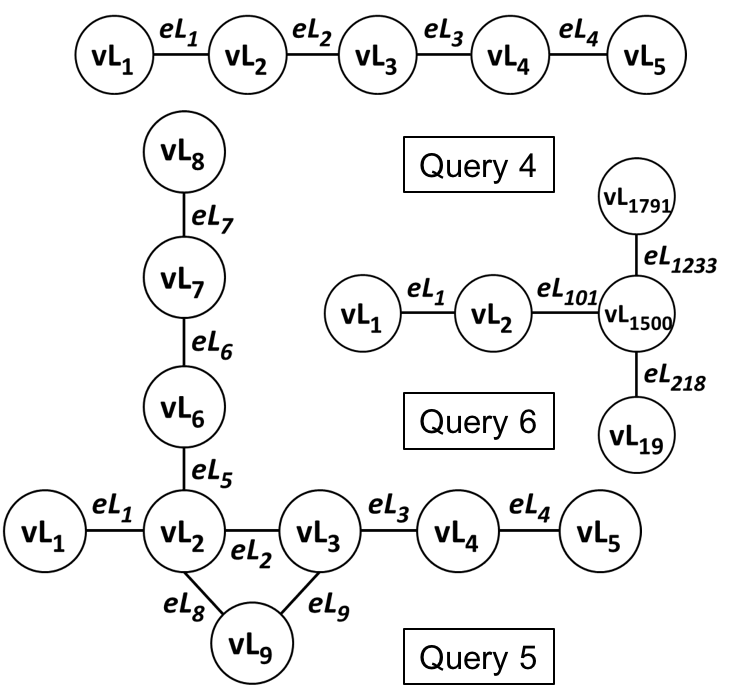}
    \caption{Queries for Synthetic Graph}
    \label{fig:SyntheticGraphQueries}
  \end{minipage}
  \hfill
  \begin{minipage}[b]{0.6\textwidth}
    \includegraphics[width=\textwidth]{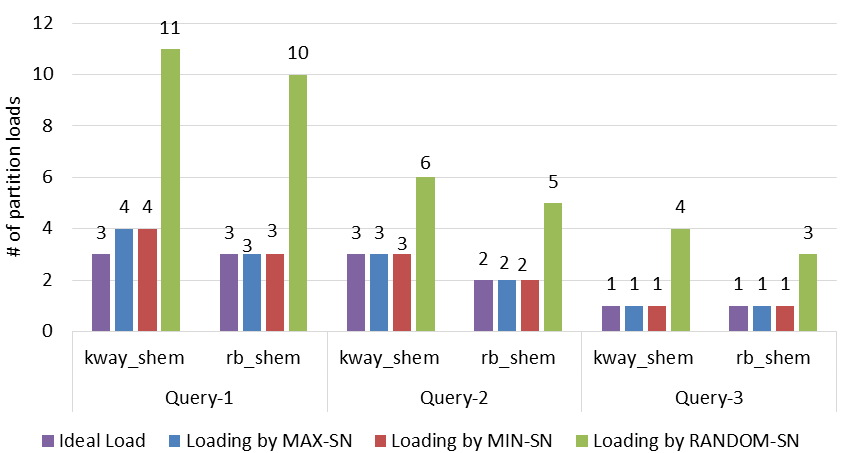}
    \caption{Performance of query answering on the IMDB graph partitioned by METIS}
    \label{fig:IMDB_METIS_MMR}
  \end{minipage}
\end{figure}

\forceindent We have used the synthetic graph mainly to test our metrics and heuristics for multiple answers spanning several partitions. Due to the synthetic nature of the graph, embedded substructures were used instead of queries that not only span partitions but also need a partition more than once. In the synthetic graph, we embed 200 instances of a substructure shown in Figure \ref{fig:SyntheticGraphQueries} that also acted as Query 5. One of the queries was a subgraph of the embedded structure (Query 4), whereas for one query \textit{only} a part (2 nodes and one edge) of the embedded substructure was present in the synthetic graph (Query 6.) Each query was executed using the MIN-SN, MAX-SN and baseline RANDOM-SN choice, on a dataset that was partitioned by 6 schemes, resulting in a total of \textit{54 experiments per dataset}.

\subsection{Evaluation of Start Node Heuristics}
\label{subsec:startinglabels}

\begin{table}[ht]
\begin{minipage}[b]{0.38\linewidth}
\centering
\begin{tabular}{|l|l|l|l|}
\hline
\multirow{2}{*}{$\textbf{h(D)}_{pschemes}^{Query_i}$} & \multicolumn{3}{|c|}{IMDB} \\
\cline{2-4}
 &
i = 1 & i = 2 & i = 3 \tabularnewline
\hline
\hline
\textbf{\textcolor{red}{MAX-SN}} & \textbf{\textcolor{red}{0.847}} & \textbf{\textcolor{red}{0.944}} & \textbf{\textcolor{red}{1.0}}  \tabularnewline
\hline
MIN-SN & 0.847 & 0.944 & 1.0\tabularnewline
\hline
RANDOM-SN & 0.302 & 0.337 & 0.320\tabularnewline
\hline
\hline
 & \multicolumn{3}{|c|}{Synthetic} \\
\cline{2-4}
 &
i = 4 & i = 5 & i = 6 \tabularnewline
\hline
\hline
\textbf{\textcolor{red}{MAX-SN}} & \textbf{\textcolor{red}{0.587}} & \textbf{\textcolor{red}{0.545}} & \textbf{\textcolor{red}{0.618}} \tabularnewline
\hline
MIN-SN & 0.385 & 0.368 & 0.401 \tabularnewline
\hline
RANDOM-SN & 0.280 & 0.260 & 0.270 \tabularnewline
\hline
\end{tabular}
\caption{Performance of SN heuristics for a single query across partitioning schemes}
\label{tab:SN1}

\end{minipage}\hfill
\begin{minipage}[b]{0.56\textwidth}
\centering
\includegraphics[width=\textwidth]{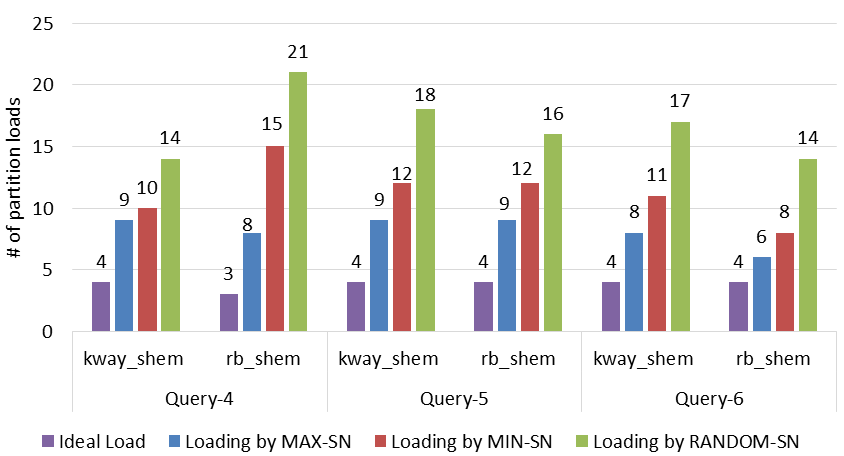}
\captionof{figure}{Performance of query answering on the Synthetic graph partitioned by METIS}
\label{fig:Synthetic_METIS_MMR}
\end{minipage}
\end{table}

\begin{table}[h]
\centering
\begin{tabular}{|l|l|l|l|l|l|l|}
\hline
\multirow{2}{*}{$\textbf{h(D)}_{\{Q1,Q2,Q3\}}^{pscheme_i}$} & \multicolumn{6}{|c|}{IMDB}\\
\cline{2-7}
 &
fast & fastsocial & eco & ecosocial & kway\_shem & rb\_shem \tabularnewline
\hline
\hline
\textbf{\textcolor{red}{MAX-SN}} & \textbf{\textcolor{red}{0.889}} & \textbf{\textcolor{red}{0.889}} & \textbf{\textcolor{red}{0.889}} & \textbf{\textcolor{red}{1.0}} & \textbf{\textcolor{red}{0.917}} & \textbf{\textcolor{red}{1.0}} \tabularnewline
\hline
MIN-SN & 0.889 & 0.889 & 0.889 & 1.0 & 0.917 & 1.0 \tabularnewline
\hline
RANDOM-SN & 0.262 & 0.317 & 0.345 & 0.306 & 0.341 & 0.344 \tabularnewline
\hline
\hline
\multirow{2}{*}{$\textbf{h(D)}_{\{Q4,Q5,Q6\}}^{pscheme_i}$} & \multicolumn{6}{|c|}{Synthetic}\\
\cline{2-7}
 &
fast & fastsocial & eco & ecosocial & kway\_shem & rb\_shem \tabularnewline
\hline
\hline
\textbf{\textcolor{red}{MAX-SN}} & \textbf{\textcolor{red}{0.613}} & \textbf{\textcolor{red}{0.571}} & \textbf{\textcolor{red}{0.603}} & \textbf{\textcolor{red}{0.756}} & \textbf{\textcolor{red}{0.463}} & \textbf{\textcolor{red}{0.495}} \tabularnewline
\hline
MIN-SN & 0.355 & 0.430 & 0.366 & 0.448 & 0.366 & 0.344 \tabularnewline
\hline
RANDOM-SN & 0.262 & 0.267 & 0.279 & 0.321 & 0.248 & 0.226 \tabularnewline
\hline
\end{tabular}
\caption{Performance of SN heuristics for a batch of queries on partitioning schemes}
\label{tab:SN2}
\end{table}

\begin{figure}[htb]
    \centering
\includegraphics[width=\textwidth]{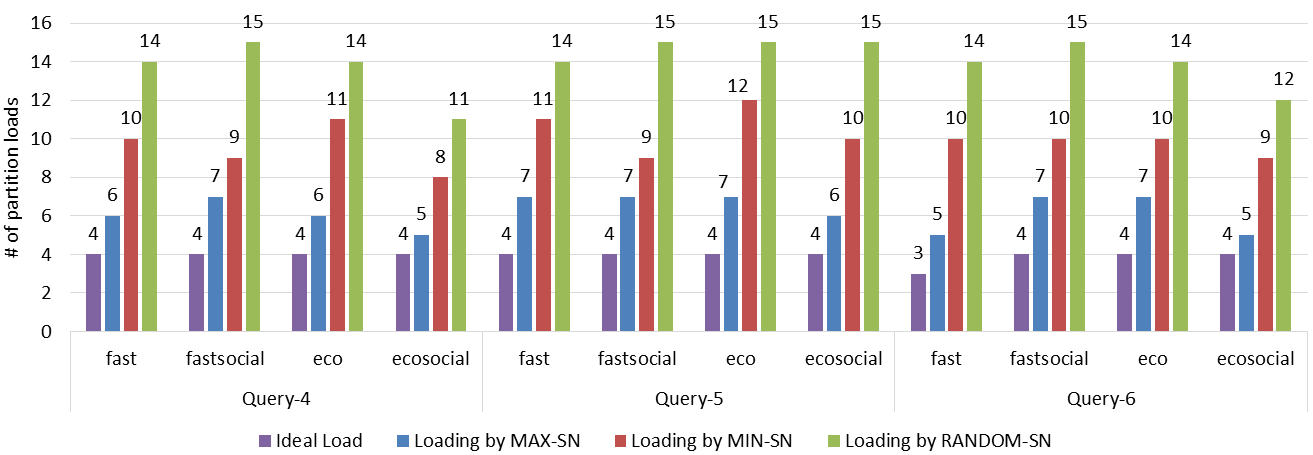}
    \caption{Performance of query answering on the Synthetic graph partitioned by KaHIP}
    \label{fig:Synthetic_KaHIP_MMR}
\end{figure}

\begin{figure}[h]
    \centering
  \includegraphics[width=\textwidth]{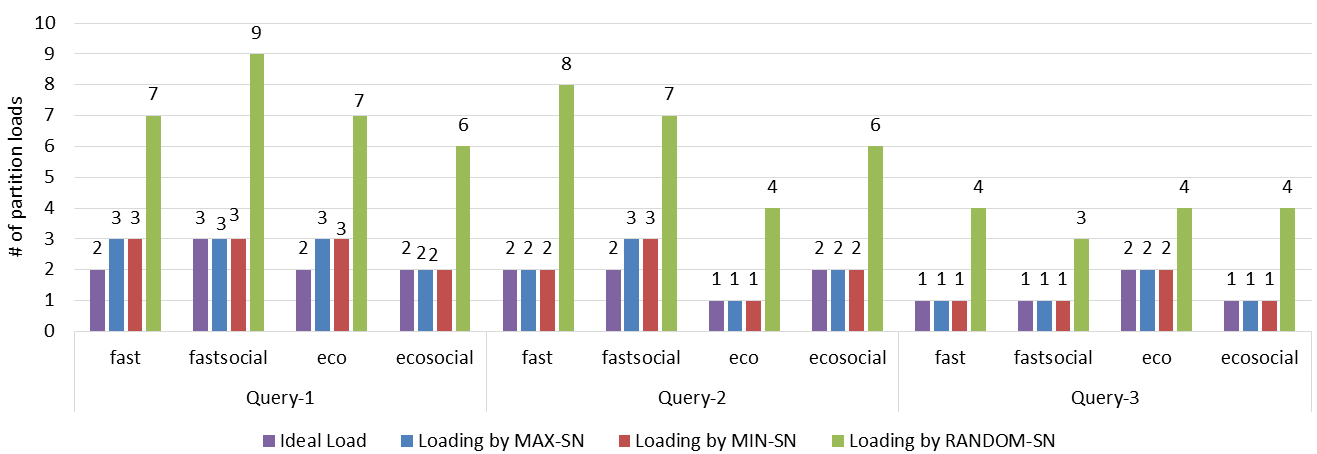}
    \caption{Performance of query answering on IMDB graph partitioned by KaHIP}
    \label{fig:IMDB_KaHIP_MMR}
\end{figure}

Figure \ref{fig:IMDB_METIS_MMR} and \ref{fig:IMDB_KaHIP_MMR} illustrate the number of partitions loaded for the ideal case, proposed heuristics (MAX-SN and MIN-SN) and the baseline (RANDOM-SN) for IMDB partitioned by METIS or KaHIP, respectively. From these figures, we obtain the values of the evaluation measures - $\textbf{h(D)}_{pschemes}^{query}$ and $\textbf{h(D)}_{qbatch}^{pscheme}$ (defined in Section \ref{sec:measures}) listed in Table \ref{tab:SN1} and Table \ref{tab:SN2}, respectively. The \textit{higher values for the proposed MIN-SN or the MAX-SN heuristic show that they have a better performance while answering single or a batch of queries on a partitioned graph as compared to the baseline RANDOM-SN choice}. Similar inference has been made for the Synthetic graph, for which the values of the measures have been calculated from Figure \ref{fig:Synthetic_METIS_MMR} and \ref{fig:Synthetic_KaHIP_MMR}.

\forceindent Further, Table \ref{tab:SN1} and \ref{tab:SN2} also show that the \textit{average load ratio for MAX-SN is greater than or equal to the average load ratio of MIN-SN heuristic, for processing a single or batch of queries across all or a single scheme, respectively}. Therefore, these experiments also validate our claim that the \textbf{MAX-SN heuristic performs as good as or better than the MIN-SN heuristic.} In case of Synthetic graph, where multiple results for each query are possible MAX-SN is always better than MIN-SN (Figure \ref{fig:Synthetic_METIS_MMR} and \ref{fig:Synthetic_KaHIP_MMR}). However, due to presence of only unique vertex labels the order and number of partition loads for IMDB queries is same for both MIN-SN and MAX-SN (\ref{fig:IMDB_METIS_MMR},  \ref{fig:IMDB_KaHIP_MMR}).

\subsection{Evaluation of Connected Components Heuristics}
\label{subsec:connectedcomponents}

The heuristic for choosing a partitioning scheme based on connected components metric is for improving performance.

\begin{table}[h]
\centering
\begin{tabular}{|l|l|l|l|l|}
\hline
\multirow{4}{*}{$\textbf{h(D)}_{\{Q1, Q2, Q3\}}^{pscheme_i}$} & \multicolumn{4}{|c|}{IMDB} \\
\cline{2-5}
 & \multicolumn{2}{|c|}{KaHIP} & \multicolumn{2}{|c|}{METIS} \tabularnewline
\cline{2-5}
 & \textbf{\textcolor{red}{MIN-CC}} & MAX-CC & \textbf{\textcolor{red}{MIN-CC}} & MAX-CC \tabularnewline
\cline{2-5}
& ecosocial & fast &  rb\_shem & kway\_shem \tabularnewline
\hline
\hline
\textbf{\textcolor{red}{MAX-SN}} & \textbf{\textcolor{red}{1.0}} & 0.889 & \textbf{\textcolor{red}{1.0}} & 0.917 \tabularnewline
\hline
MIN-SN & 1.0 & 0.889 & 1.0 & 0.917 \tabularnewline
\hline
\# Total CC & 40975 & 77687 & 58371 & 80417 \tabularnewline
\hline
\hline
\multirow{4}{*}{$\textbf{h(D)}_{\{Q4, Q5, Q6\}}^{pscheme_i}$} & \multicolumn{4}{|c|}{Synthetic} \\
\cline{2-5}
 & \multicolumn{2}{|c|}{KaHIP} & \multicolumn{2}{|c|}{METIS} \tabularnewline
\cline{2-5}
 & \textbf{\textcolor{red}{MIN-CC}} & MAX-CC & \textbf{\textcolor{red}{MIN-CC}} & MAX-CC \tabularnewline
\cline{2-5}
& ecosocial & fast &  kway\_shem & rb\_shem \tabularnewline
\hline
\hline
\textbf{\textcolor{red}{MAX-SN}} & \textbf{\textcolor{red}{0.756}} & 0.613 & \textbf{\textcolor{red}{0.463}} & 0.495 \tabularnewline
\hline
MIN-SN & 0.448 & 0.355 & 0.366 & 0.344 \tabularnewline
\hline
\#Total CC & 5367 & 14365 & 16606 & 17316 \tabularnewline
\hline
\end{tabular}
\caption{Performance of CC heuristics on a Batch of Queries}
\label{tab:CC}
\end{table}

\forceindent Table \ref{tab:CC} shows the evaluation of the partitioning schemes that generated the highest and least number of total connected components by measuring their performance for a batch of queries ($\textbf{h(D)}_{qbatch}^{pscheme}$). In most of the scenarios, \textit{the higher values for the schemes that generate the least number of total connected components, irrespective of the MAX-SN or MIN-SN heuristic, validate our claim that choosing a scheme using MIN-CC heuristic serves as better choice for answering queries on partitioned graphs as compared to MAX-CC heuristic}.

However, if the difference between highest and the least number of total connected components is not significant then no definite choice can be made. This case is exemplified by the MET-IS partitioned Synthetic graph in Table~\ref{tab:CC} where both the listed schemes have similar performance because the difference between the number of total connected components produced by them falls below $5\%$. Therefore, the ecosocial partitioning scheme that always generates \textit{substantially low number of total connected components} as compared to the other three schemes, becomes the preferred partitioning scheme using MIN-CC heuristic.

\forceindent Thus, section \ref{subsec:startinglabels} and \ref{subsec:connectedcomponents}, empirically validate that \textbf{MAX-SN} \textbf{and} \textbf{MIN-CC} independently are better than other heuristics. Furthermore, they can be combined to improve the performance even more as can be seen from Table \ref{tab:CC}. 
\section{Parallel Partition Processing Using TraditionalMP}
\label{sec:approach-2}




The OPAT approach described in detail and evaluated in the previous sections concentrated on scalability based on partitions, and its correctness. Loading one partition at a time is  the easiest way to achieve scalability with very limited resources.  The TraditionalMP and MapReduceMP approaches, discussed in this and the next section, address minimizing the response time by leveraging parallel processing on partitions to the extent possible. We explore the traditional as well as the map/reduce multi-processor  approaches as there are significant differences in how the query processing algorithm is designed and partial results created and merged. It is easier to understand the TraditionalMP approach as it is an extension of the OPAT approach with a focus on determining how many partitions to load in each iteration for processing them in parallel based on resource availability.

\forceindent Although the total number of partitions needed in any round can not be  known \textit{a priori} (or determined \textit{a priori}), we need to consider two cases for analysis: i) enough processors are available to process \textit{all the partitions} in parallel in any given iteration $i$ (termed as \textit{required($i$)}) and ii) only \textit{$p$} processors are available for any iteration which is less than \textit{required($i$)}. In fact, \textit{$p$} as 1 is the OPAT approach. Note that the following observations do not depend on the total number of partitions of the graph. However, if the number of partitions is greater, it is likely that the number of processors needed in many iterations is likely to increase, but the computation time used in each round is likely to decrease (due to smaller partition sizes.) Also, the number of iterations needed for answering a query is likely to increase if the number of graph partitions increase due to increase in the likelihood of answers spanning more than one partition. Increasing the number of partitions may also positively benefit query processing as the number of connected components in each partition is likely to reduce. \textit{Clearly, there is a trade off between the number of processors available, number of partitions of the graph, and the number of partitions that are needed  in each iteration for answering a query (or a batch of queries.)}

\subsection{TraditionalMP Algorithm}

\begin{algorithm}[!htb]
			\caption{Traditional Parallel Query Processing Algorithm (TraditionalMP)}
			\label{alg:tradMP}
			\begin{algorithmic}[1]
				\REQUIRE $n$ partitions, $p$ processors, a query plan $qp$, and  a heuristic $h$ to use
				\ENSURE All answers to the query $q$ using the query plan $qp$
                \STATE \quad Initialize SNI (Starting Node Information) file (\textbf{see Section~\ref{sec:metric1}})
				\STATE \quad EP = Identify all eligible partitions from the SNI file 
				\STATE \quad Initialize $p$ InterMediate Answers (IMA) File, one for each partition
				\STATE \quad CP = choose $p$ partitions from EP using heuristics h \textbf{//chosen partitions}
				\WHILE {CP is not empty}
				\STATE \quad Assign partitions to processors
				\STATE \quad Execute the PGQP algorithm in each processor
				\STATE \quad IMA$_i$ is generated by the $i^{th}$ ~processor
				\STATE \quad Merge IMA$_i$ files into FAA file (\textbf{see Section~\ref{sec:metric1}})
				\STATE \quad Recompute metrics based on this iteration
				\STATE \quad update the SNI file using IMA files (\textbf{see Section~\ref{sec:metric1}})
				\STATE   \quad update EP file from the SNI file
				\STATE \quad CP = choose $p$ partitions from EP using heuristics h \textbf{//chosen partitions}
				\ENDWHILE
				\STATE \quad FAA file contains all the answers
			\end{algorithmic}
			\end{algorithm}

Algorithm~\ref{alg:tradMP} uses the same extended partition graphs as input and code used by the OPAT approach (i.e., PGQP) and generates the same intermediate answers files. Merging of intermediate answers (IMA) files into final all answers (FAA) file is done at the end of each iteration. The merging can be done one file at a time as is done in the OPAT approach and the order does not matter.  The SNI file is also updated after each iteration. The main difference between the OPAT approach and this approach is in the set of   partitions chosen for each iteration (CP) and allocation of processors to partitions. When there are no more eligible partitions to be processed, the algorithm stops yielding all answers in FAA file.

\forceindent There is a difference between the two cases where the available processors are  $p$ and required($i$ in iteration $i$). If the number of available processors are \textit{required($i$)}, all eligible partitions can be processed in parallel in each iteration. Otherwise, some heuristics need to be used for choosing $p$ partitions out of eligible partitions. Note that the number of eligible partitions is not the same in each iteration. It depends on the query and the partition characteristics and hence can only be determined at run time. Lines 2 and 12 update the eligible partitions based on the information in SNI file. It is possible that some paths will terminate as they do not satisfy query constraints. Other paths will continue in other partitions and this information is added to the SNI file at the end of each iteration.

\forceindent Line 4 and 13 determine partitions chosen for the current iteration (CP) using EP and h discussed in Section~\ref{sec:metric1}. In case of $ p ~ < required(i)$, the partitions can be ordered using  either MIN-SN or MAX-SN heuristic and top $p$ partitions are chosen. This is moot for the RANDOM-SN choice. The final all answers file (FAA) is updated after each iteration using all intermediate answers files (the order doesn't matter.)

\subsection{Heuristics and Response Time Discussion for  TraditionalMP}

If required number of processors are available for each iteration, a good estimate for the upper bound on the number of iterations required is the maximum path length of the query plan (which is a tree.) The actual or average case is likely to be much smaller because, in each iteration, more than one edge of the query plan is likely to be traversed which will reduce the total number of iterations. This is a significant departure from the OPAT approach. However, for the case where the number of processors available is less than \textit{required($i$)} in iteration $i$, the number of iterations may increase gradually reaching the OPAT case where $p$ is one.

\forceindent In fact, $p$ being less than \textit{required($i$)} may not be as pronounced if some of the answers span into partitions that are \textit{yet to be processed}. Our heuristics take this into account by recomputing the number of start nodes in each partition after each iteration. Analytically, it is not possible to predict this as it depends on query plan structure, conditions in the query plan that have to be computed at run time, and graph characteristics. As for the number of connected components heuristic is concerned (MIN-CC, MAX-CC, and RANDOM-CC), there should not be any difference in terms of their effect between the OPAT approach and TraditionalMP approach as both work in the same way.

\forceindent From a response time perspective, TraditionaldMP approach will certainly improve the response time significantly for any $p$ greater than one. The amount of improvement depends on several factors: i) matching of processor capability with partition size as the iteration time is determined by the longest time used by a processor, ii) number of processors available as it \textit{may change} the number of iterations required,  iii) partition and query characteristics in terms of the distance of starting nodes to the end of the partition, iv)  maximum number of partitions spanned by an answer, and v) importantly, whether the same partition is needed more than once. Some of these can be empirically measured by a set of well-chosen experiments. Due to space constraints, we have not been able to include experimental results in this~\pubtype. If the number of processors available is slightly less than \textit{required($i$)}, that may not increase the response time significantly. It will be interesting to analyze the effect of $p$ and its effect on response time as it changes from  1 to $max(required($i$))$. If an answer spans multiple connected components \textit{within a partition}, that necessitates loading that partition as many times even in the parallel processing approach. This also indicates why MIN-CC has come out as a  better heuristic than others (\textbf{see Section~\ref{subsec:connectedcomponents}}.)

\forceindent A much greater response time benefit can be obtained when a batch of queries are processed using this approach instead of a single query. However, for this the book keeping complexity increases as one needs to keep track of not only different answers for the same query but also separate answers for each query correctly. We are investigating this as future work.

\section{Parallel Partition Processing Using MapReduceMP}
\label{sec:approach-3}





Use of Map/Reduce paradigm is another promising alternative for processing partitions independently in parallel and assemble the answers correctly. Anyone familiar with map/reduce paradigm will immediately discern that the algorithm used in OPAT (and TraditionalMP) can not be \textit{directly} used either in the map function or in the reducer. The requirements of a map function to process each input record independently and emit a key/value entails that we view this problem and computation differently. Preferably, the reducer should be able to assemble intermediate results emitted by the mapper and grouped the way we have done in the previous approaches and update the eligible partitions set (EP) for the next iteration and  FAA using IMA files as well as the SNI file for the next iteration. For this approach, we lean on one of earlier work where we have mapped a substructure discovery algorithm into the map/reduce framework (for scalability and speed up) and have conducted extensive experiments and cost analysis~\cite{tkde/DasC18,phdThesis/Das17}.

\forceindent Unlike the previous two approaches, each query plan node is expanded by one edge on multiple eligible partitions in each iteration by the mapper task. In the previous approaches, we processed all answer paths \textit{within a partition} in each iteration. This cannot be done in this approach as we cannot assume the input adjacency list of a partition to be presented in some order of nodes. Actually, there may not even be an order that allows us to compute all paths within a partition in the same map/reduce computation. Since we are expanding one edge at a time in this approach, the minimum number of iterations required will correspond to the maximum query plan path length.

\subsection{MapReduceMP Algorithm}
Each graph partition along with cut set information is used as input to the mapper in the form of an adjacency list. Expansion occurs either on a vertex id or on a label as we do in the previous approaches. To facilitate expansion on the node label, the adjacency list used for substructure discovery~\cite{dawak/DasC15} is slightly modified to include vertex label in addition to vertex id as part of the key. Hence, the input to the mapper has $<$ vertex id, label$>$ as its key and the adjacency list as value. The adjacency list will include partition id for each node as well. This will allow us to identify when an edge expansion goes out of the current partition indicating that answer evaluation needs to be continued in a different partition indicated by the partition number associated with the node. In addition to the input, the same SNI file is used by all mapper tasks and the partition-relevant IMA file is used by each partition.

\forceindent Although the query plan contains only labels, once the query plan evaluation starts, the expansion will lead to nodes whose ids are known. Hence, it should be possible to expand either on a node label or using a node id during any iteration. This is done either by using the node id or the node label of the key  using the SNI file.
In all iterations, the mapper uses the SNI file to find the vertex ids or the label and expands the vertices by adding an edge from the adjacency list. Conditions present in the query
are also applied to the edge labels and vertex labels to determine whether the answer evaluation will continue further. For each one-edge expansion, the mapper emits the partition id in which the next expansion needs to takes place as the key and the the node id and its label of start and end nodes as the value. The expansion could be either in the \textit{same partition or in a different partition.}  Note that if continuation is in a different partition that information comes out correctly along with the node id and label of the node in the output emitted by the mapper.
Shuffle groups the output of mappers on the partition id and sends them to corresponding reducers. A combiner can also be used to improve response time. Since the amount of work done in a reducer is not as much as a mapper, less than \textit{required($i$)} reducers can be used without sacrificing response time.

\begin{figure}[htb]
    \centering
    \includegraphics[width=0.95\textwidth]{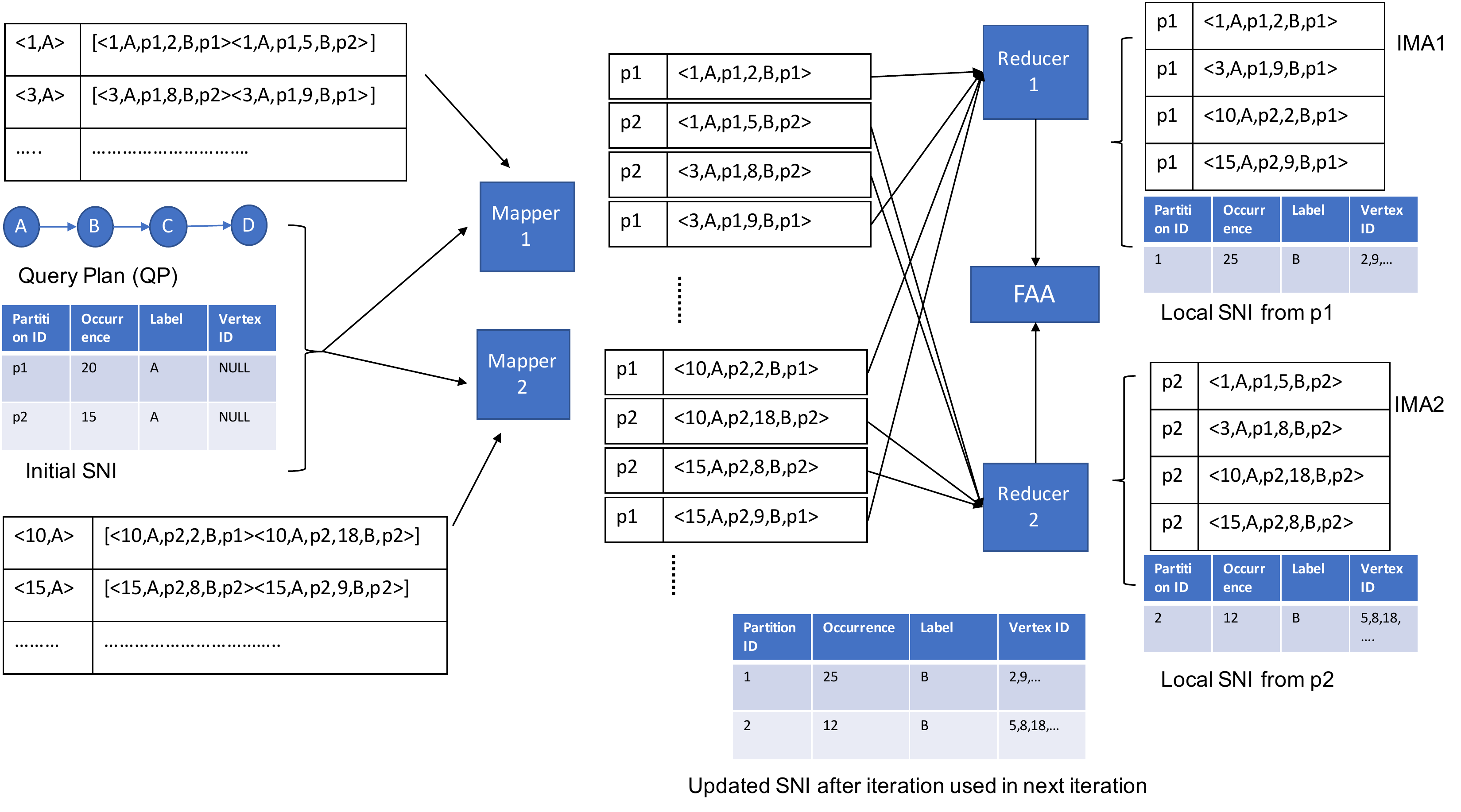}
    \vspace{-10pt}
    \caption{First iteration for query processing using Map/Reduce}
    \label{fig:workflow_mrMP}
\end{figure}

\forceindent The reducer processes key value lists generated for each partition. The reducer does the following by traversing the value list for each partition id once: i) updates the SNI file by removing the nodes expanded in this iteration (whether on label or node id). Also, new nodes and their partition ids are inserted into the SNI file for the next iteration, ii) IMA file for that partition is updated using the edge information for each answer. Since both start and end node ids are present, it is not difficult to separate answers based on their path within and across the partitions, and iii) finally, the FAA file is updated with partial or complete answers while processing the value list. The iterations end when the SNI file is empty leaving all answers in the FAA file. Fig.~\ref{fig:workflow_mrMP} shows the first iteration of MapReduceMP algorithm. The map/reduce driver (or jobtracker) is responsible for adding the local SNIs across reducer to create the updated SNI file for use in next iteration. Final answers emitted by each reduce task is also merged by the driver into a single FAA file (order of answers does not matter.)

\subsection{Heuristics and Response Time Discussion for MapReduceMP}
The correctness of the algorithm follows from what is generated in each iteration and how they are used for updating SNI, IMA, and FAA files. Instead of a path, a subset of a path is generated in each iteration. All the information emitted for continuing the query evaluation is the same as in the previous approaches but different algorithmically.

\forceindent Given $p$ partitions we will have \textit{required($i$)} \textit{mapper tasks} in iteration $i$. If we have $m$ mapper nodes, we can either assign $m$ mapper tasks in each iteration or \textit{required($i$)} mapper tasks in each iteration even if $m ~<$ \textit{required($i$)}. In this case, multiple mapper tasks will be processed by the same mapper node. Note that \textit{required($i$)} may be different in each iteration. For the case  $m < maximum(required(i))$, allocation of $m$ mapper tasks for that iteration is better than allocating \textit{required($i$)} mapper tasks. This is because the completion of an iteration (and hence response time) is dictated by the slowest mapper node completion. On the other hand, since the number of partitions eligible for execution in each iteration (i.e., \textit{required($i$)}) varies in each iteration, the mapper tasks may get amortized over the number of iterations. However, if $m << maximum(required(i))$, the number of iterations will increase
there by increasing the response time, hopefully asymptotically reaching OPAT when $m$ is one. There is a need to study the trade off between allocation of more than one mapper task to a mapper node versus increasing the number of iterations from a response time perspective. Certainly, the best response time is obtained when the number of mappers is the same as \textit{required($i$)} for every iteration. In contrast, the number of reducers needed is small (compared to \textit{required($i$)}) as the amount of work done in updating SNI, IMA and FAA files is quite negligible as it depends on the number of answers being computed and not on the size of any partition being processed as is the case for a mapper task.

The heuristics of MAX-SN and MIN-SN can be used for choosing the partitions to be processed in any iteration when $m < required(i)$ as in the previous approaches. Similarly,the number of connected components heuristic (MIN-CC, MAX-CC, and RANDOM-CC) can also be used in the same way. The usage of heuristics should produce the same performance trend in all the three approaches.

\subsection{Desiderata}
Development of the OPAT approach was critical for establishing scalability and correctness of partitioned query processing. Using that, we have extended the work to improve response time based on the availability of resources. The motivation for the map/reduce approach comes from its ability to provide resources on demand and match partition sizes based on that. Both approaches support varying number of processors with response time trade off. The algorithms are different although inputs are same modulo representation. We believe that further improvements are possible for the map/reduce approach using combiners and applying component cost analysis as we have done for the substructure discovery approach~\cite{phdThesis/Das17}. We also believe that both traditional and map/reduce approaches are needed in different contexts for achieving scalability and response time improvement. The suite of algorithms presented in this ~\pubtype~ provides alternatives for response time trad off. More importantly, all approaches are independent of the partitioning strategy so the user can use different partitioning strategies based on query and graph characteristics.

\section{Conclusions}
\label{sec:conclusions}

\forceindent In this~\pubtype, we have proposed alternative approaches for processing queries over partitions of a large graph database. These approaches range from processing one partition at a time (termed OPAT) to processing multiple partitions in parallel using the traditional (termed TraditionalMP) and currently popular map/reduce framework (termed MapReduceMP). Each has its own strengths and weaknesses which are analyzed and compared. The goal of this work is to move towards efficient and scalable query processing techniques for large graphs.

\forceindent Partitioning is one way to ensure scalability of query processing on a graph database of any size. Parallel processing further helps improve the response time critical for i) large graph sizes and ii) large number of partitions. We have addressed correctness and proposed heuristics to reduce the amount of work done (in terms of the number of partitions loaded) for query processing. Implementation of one of the proposed approaches as well as validation of the proposed heuristics for all approaches has been shown on real-world and synthetic data sets.  

\forceindent Beyond this, we are exploring the use of parallel processing of partitions, heuristics for optimizing the work done for a \textit{batch of queries}, and partitioning strategies to improve the efficiency of  query processing.
\bibliographystyle{plain}
\bibliography{./bibliography/somu_research,./bibliography/itlabPublications,./bibliography/itlabTheses,./bibliography/graph-search,./bibliography/graph-partitioning,./bibliography/JayBib}
\end{document}